\documentclass[article]{setup}
\usepackage{
  thumbpdf,
  lmodern,
  bm,
  amsmath,
  tikz,
  subcaption,
  algorithm,
  algpseudocode,
  dcolumn,
  booktabs,
  siunitx,
  multirow,
  makecell,
  xspace,
  multicol
}

\usepackage[utf8]{inputenc}
\usetikzlibrary{math}
\usetikzlibrary{arrows.meta}

\newcommand{\gRbase}{\pkg{gRbase}\xspace}
\newcommand{\grbase}{\pkg{gRbase}\xspace}
\newcommand{\gRain}{\pkg{gRain}\xspace}
\newcommand{\bayesnet}{\pkg{BayesNetBP}\xspace}
\newcommand{\sparta}{\pkg{sparta}\xspace}
\newcommand{\jti}{\pkg{jti}\xspace}
\newcommand{\ess}{\pkg{ess}\xspace}

\author{Mads Lindskou\\ Aalborg University \And Søren Højsgaard\\Aalborg University \AND Poul Svante Eriksen\\ Aalborg University \And Torben Tvedebrink \\ Aalborg University} 
\Plainauthor{Mads Lindskou, Søren Højsgaard, Poul Svante Eriksen, Torben Tvedebrink}

\title{sparta: Sparse Tables and their Algebra with a View Towards High Dimensional Graphical Models}

\Plaintitle{sparta: Sparse Tables and their Algebra for use in High Dimensional Graphical Models}
\Shorttitle{\sparta: Sparse Tables}
    
\Abstract{A graphical model is a multivariate (potentially very high
  dimensional) probabilistic model, which is formed by combining
  lower dimensional components. Inference (computation of conditional
  probabilities) is based on message passing algorithms that utilize
  conditional independence structures. In graphical models for
  discrete variables with finite state spaces, there is a fundamental
  problem in high dimensions: A discrete distribution is represented
  by a table of values, and in high dimensions such tables can become
  prohibitively large. In inference, such tables must be multiplied
  which can lead to even larger tables. The \sparta package
  meets this challenge by implementing methods that efficiently
  handles multiplication and marginalization of sparse tables. The
  package was written in the \proglang{R} programming language and is
  freely available from the Comprehensive \proglang{R} Archive Network
  (CRAN). The companion package \jti, also on CRAN, was developed to
  showcase the potential of \sparta in connection to the Junction Tree
  Algorithm. We show, that \jti is able to handle highly complex graphical models which are otherwise
  infeasible due to lack of computer memory, using \sparta as a backend for table operations.
}

\Plainkeywords{
  sparse table, 
  Bayesian network, 
  graphical model, 
  R, 
  C++, 
  \sparta, 
  \jti}

\Address{
  Mads Lindskou\\
  Department of Mathematical Sciences\\
  Aalborg University\\
  Skjernvej 4A, DK-9220 Aalborg {\O}st\\
  E-mail: \email{mads@math.aau.dk}\\
  URL: \url{https://github.com/mlindsk}
}

\begin{document}


\section[Introduction]{Introduction}
\label{sec:intro}

A multivariate probability distribution for discrete variables with finite state
spaces can be represented by a multi-dimensional array.  However, for
high-dimensional distributions where each variable may have a large
state space, lack of computer memory can become a problem. For
example, an $80$-dimensional random vector in which each variable
has $10$ levels will lead to a state space with $10^{80}$ cells. Such
a distribution can not be stored in a computer; in fact, $10^{80}$ is
one of the estimates of the number of atoms in the universe. However,
if the array consists of only a few non-zero values, we need only store
these values along with information about their location. That is, a sparse representation of a table. We describe here the \proglang{R} \citep{R} package \sparta \citep{sparta} for efficient multiplication and marginalization of sparse tables.

The family of graphical models \citep{pearl2014probabilistic, lauritzen1996graphical, hojsgaard2012graphical} is vast, and includes many different models. In this paper, we consider Bayesian networks and decomposable undirected graphical models. Undirected graphical models are also known as Markov random fields (MRFs). Decomposability is a property ensuring a closed form of the maximum likelihood parameters. Graphical models, enjoy the property that conditional independencies can be read off from a
graph consisting of nodes, representing the random variables. In Bayesian networks, edges are directed from a node to another and represents a directed connection. In a MRF the edges are undirected and these should be regarded as associations between pairs of nodes in a broad sense. A decomposable MRF can be turned into a Bayesian network while retaining the correct likelihood function. However, one can no longer interpret conditional independencies in this transformed graph. This transformation is merely a step towards making inference in decomposable MRFs and as such, we shall mainly focus on introducing concepts regarding Bayesian networks.

We illustrate \sparta using the \jti package \citep{jti} which implements the Junction Tree Algorithm (JTA) for discrete variables using the Lauritzen-Spiegelhalter updating scheme \citep{lauritzen1988local} with table operations relying on \sparta. JTA is used for inference in Bayesian networks (BNs). We also show, that multiplication and marginalization of sparse tables is fast compared with standard built-in \proglang{R} functions and is comparable to \gRbase when the tables are sparse.

In addition to \jti, there are to our knowledge three other packages
for belief propagation in \proglang{R}; \pkg{gRain} \citep{grain},
\bayesnet \citep{BayesNetBP} and \pkg{RHugin} \citep{rhugin} where the
latter is not on the Comprehensive \pkg{R} Archive Network (CRAN). The
only \pkg{R} package on CRAN that has an API for (dense) table
operations is \gRbase, which \pkg{gRain} depends upon. Some \pkg{R}
packages that rely on \pkg{gRain} and \gRbase are \pkg{geneNetBP}
\citep{genenetbp} and \pkg{bnspatial} \citep{bnspatial}. The
\pkg{bnclassify} package \citep{bnclassify} has a lower \proglang{C++}
class implementation of conditional probability tables. Our goal is
twofold: Firstly, to provide an efficient back-end for sparse table
operations for other \pkg{R} packages. Secondly, to provide a new implementation of JTA, namely \jti, using \sparta as a back-end.

In Section \ref{sec:notation} we introduce basic notation and terminology and in Section \ref{sec:messagepassing} we motivate the usage of sparse tables through JTA. Section \ref{sec:intui} serves as a primer to our novel representation of tables and their algebra given in Section \ref{sec:sparse_tables}. In Section \ref{sec:sparse_tables} we also demonstrate how to use \sparta. Section \ref{sec:usecases} outlines how to use \jti and gives specific examples, using two BNs which are well known from the literature. Moreover, we demonstrate the strength of \sparta using the \ess \citep{ess} package to fit a decomposable MRF on real data. In Section \ref{sec:benchmark} we show, that the trade-off between execution time and memory allocation using \sparta is acceptable for small and medium sized tables, and comparable to \grbase in high dimensional sparse tables.

Finally, we mention that \sparta leverages from the \pkg{RcppArmadillo} package \citep{rcpparmadillo} by implementing compute-intensive procedures in \proglang{C++} for better run-time performance.

\section{Notation and Terminology}
\label{sec:notation}
Let $p$ be a discrete probability distribution for the random vector $X = X_{V} = (X_{v} \mid v \in V)$ where $V$ is a set of labels of the random variables $X_{v}, v\in V$ and where each $X_{v}$ has finite state space. A realized value $x = (x_{v})_{v\in V}$ is called a \textit{cell}. If we denote by $I_{v}$ the finite state space of variable $X_{v}$, the domain of $p$ is given by
\[
  I = \times_{v\in V} I_{v}.
\]
Given a subset $A$ of $V$, the \textit{A-marginal cell} of $x_{V}$ is the vector, $x_{A} = (x_v)_{v\in A}$, with state space $I_{A} = \times_{v\in A} I_{v}$. A \textit{Bayesian Network} can be defined as a \textit{directed acyclic graph} (DAG), see Figure \ref{fig:graphs}, for which each node represents a random variable together with a joint probability of the form
\begin{align}
  p(x) = \prod_{v\in V} p(x_{v} \mid x_{pa(v)}),
\end{align}
where $x_{pa(v)}$ denotes the \textit{parents} of $x_v$; i.e.\ the set of nodes with an arrow pointing towards $x_v$ in the DAG. Also, $x_v$ is said to be a \textit{child} of the variables $x_{pa(v)}$. Notice, that $p(x_{v} \mid x_{pa(v)})$ has domain $I_{v} \times I_{pa(v)}$. Hence, we can encode the conditional probabilities in a table, say $\phi(x_{v}, x_{pa(v)})$, of dimension $|I_{v}|\cdot |I_{pa(v)}|$. It is also common in the literature to refer to these tables as \textit{potentials} and we shall use these terms interchangeably. In general, a potential does not have an interpretation. Sometimes, we also use subscript notation to explicitly show the set of variables for which a potential depends on. That is, $\phi_{A}$ is a potential defined over the variables $X_{A}$. The product, $\phi_{A} \otimes \phi_{B}$, of two generic tables over $A$ and $B$ is defined cell-wise as
\[
  (\phi_{A} \otimes \phi_{B})(x_{A\cup B}) := \phi_{A}(x_{A})\phi_{B}(x_{B}).
\]
In other words, the product is defined over the union of the variables of each of the two potentials. Division of two tables, $\phi_{A} \oslash \phi_{B}$, is defined analogously. The marginal table, $\phi_{A}^{\downarrow B}$, over the variables $B\subseteq A$ is defined cell-wise as
\[
  \phi_{A}^{\downarrow B}(x_B) := \sum_{x_{A\setminus B} \in I_{A\setminus B}}\phi_{A}(x_{B}, x_{A\setminus{B}}).
\]
Finally, for some $B \subseteq A$, fix $x^{\ast}_{B}$. The $x^{\ast}_{B}$ slice of $\phi_A(x)$ is then given by
\[
  \phi_{A}^{x^{\ast}_B}(x_{A\setminus B}) = \phi_{A}(x_{A\setminus B}, x^{\ast}_{B}).
\]

\section{Motivation Through Message Passing in Bayesian Networks}
\label{sec:messagepassing}
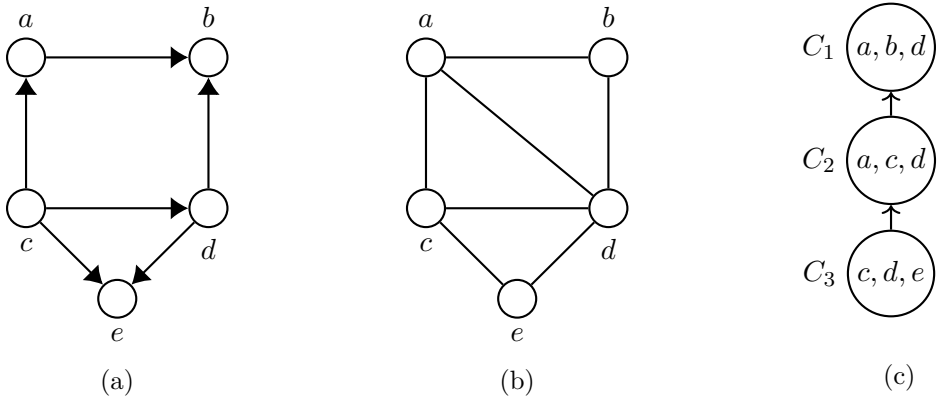
\begin{figure}[h!]
  \begin{minipage}[h]{0.33\linewidth}
    \centering
    \begin{tikzpicture}
      \tikzmath{\eps = 0.8;}
      \tikzmath{\sep_ = .5cm;}
      \tikzset{cir/.style={thick, circle, draw = black, minimum size = \sep_, inner sep = 0}}

      \node [cir, label = above:{$a$}] (a) at (-1.5*\eps, \eps){};
      \node [cir, label = above:{$b$}] (b) at (1.5*\eps, \eps){};
      \node [cir, label = below:{$c$}] (c) at (-1.5*\eps, -1.5*\eps){};
      \node [cir, label = below:{$d$}] (d) at (1.5*\eps, -1.5*\eps){};
      \node [cir, label = below:{$e$}] (e) at (0*\eps,  -3*\eps){};

      \draw [-{Latex[width=3mm]}, thick, draw=black] (a) -- (b);
      \draw [-{Latex[width=3mm]}, thick, draw=black] (c) -- (a);
      \draw [-{Latex[width=3mm]}, thick, draw=black] (c) -- (d);
      \draw [-{Latex[width=3mm]}, thick, draw=black] (c) -- (e);
      \draw [-{Latex[width=3mm]}, thick, draw=black] (d) -- (b);
      \draw [-{Latex[width=3mm]}, thick, draw=black] (d) -- (e);
    \end{tikzpicture}
    \subcaption{\label{tikz:a}}
  \end{minipage}
  \begin{minipage}[h]{0.33\linewidth}
    \centering
    \begin{tikzpicture}
      \tikzmath{\eps = 0.8;}
      \tikzmath{\sep_ = .5cm;}
      \tikzset{cir/.style={thick, circle, draw = black, minimum size = \sep_, inner sep = 0}}

      \node [cir, label = above:{$a$}] (a) at (-1.5*\eps, \eps){};
      \node [cir, label = above:{$b$}] (b) at (1.5*\eps, \eps){};
      \node [cir, label = below:{$c$}] (c) at (-1.5*\eps, -1.5*\eps){};
      \node [cir, label = below:{$d$}] (d) at (1.5*\eps, -1.5*\eps){};
      \node [cir, label = below:{$e$}] (e) at (0*\eps,  -3*\eps){};

      \draw [thick, draw=black] (a) -- (b);
      \draw [thick, draw=black] (a) -- (d);
      \draw [thick, draw=black] (c) -- (a);
      \draw [thick, draw=black] (c) -- (d);
      \draw [thick, draw=black] (c) -- (e);
      \draw [thick, draw=black] (d) -- (b);
      \draw [thick, draw=black] (d) -- (e);
      
    \end{tikzpicture}
    \subcaption{\label{tikz:b}}
  \end{minipage}
  \begin{minipage}[h]{0.3\linewidth}
    \centering
    \raisebox{-0em}{\begin{tikzpicture}
      \tikzmath{\eps = 1;}
      \tikzmath{\sep_ = .7cm;}
      \tikzset{dot/.style={thick, circle, draw = black, minimum size = \sep_, inner sep = 0, fill = black}}
      \tikzset{cir/.style={thick, circle, draw = black, minimum size = \sep_, inner sep = 2}}

      \node [cir, label = left:{$C_1$}] (C1) at (0*\eps, -1.5*\eps){$a,b,d$};
      \node [cir, label = left:{$C_2$}] (C2) at (0*\eps, -3*\eps){$a,c,d$};
      \node [cir, label = left:{$C_3$}] (C3) at (0*\eps, -4.5*\eps){$c,d,e$};
      
      \draw [thick, draw=black, ->] (C3) -- (C2);
      \draw [thick, draw=black, ->] (C2) -- (C1);
      
    \end{tikzpicture}}
    \newline
    \vspace{-0.1cm}
    \subcaption{\label{tikz:c}}
  \end{minipage}
  \caption{\label{fig:graphs}(a) A DAG. (b) A moralized and triangulated version of (a). (c) A rooted junction tree representation of (b) with root $C_1$.}
\end{figure}

Consider the simple DAG given in Figure \ref{tikz:a}, from which the joint density can be read off:
\begin{align}\label{eq:pot_rep}
  p(x_a, x_b, x_c, x_d, x_e) &= p(x_c)p(x_a \mid x_c)p(x_b \mid x_a, x_d)p(x_d \mid x_c)p(x_e \mid x_c, x_d)
\end{align}

If, for example, interest is in the joint distribution of $(x_a, x_d)$ we have to sum over $x_b, x_c$ and $x_e$ and exploiting the factorization we could calculate this as
\begin{align*}
p(x_a, x_d) = \sum_{x_c} p(x_c) p(x_a \mid x_c) p(x_d \mid x_c)\sum_{x_e}p(x_e \mid x_c, x_d)\sum_{x_b}p(x_b \mid x_a, x_d).
\end{align*}

The \textit{Junction Tree Algorithm (JTA)} can be seen as an algorithm for automatically factorizing in order to circumvent the direct summation as described in what follows using a minimal example \citep[a more general and technical exposition of the algorithm can be found in e.g.][]{hojsgaard2012graphical}: First

\begin{itemize}
\item \textit{moralize} the DAG; i.e.\ connect nodes that share a common child node,
\item remove directions in the DAG to obtain an undirected graph and
\item \textit{triangulate} the resulting graph.
\end{itemize}

Moralization ensures that the corresponding parent and child nodes are put in the same \textit{maximal clique}. A clique is a subset of the nodes for which the induced subgraph is complete and it is maximal if it is not contained in any other clique. From here, by clique, we always mean a maximal clique and we refer to these as the cliques of the graph.

A graph is triangulated if it has no \textit{cycles} of length greater than $3$. If such cycles are present, we must add \textit{fill edges} to produce a triangulated graph. A triangulated graph is also called \textit{decomposable}, and hence the connection to decomposable MRFs shows here. Finding an optimal triangulation (in terms of minimizing the number of fill edges) is a NP hard problem, but there exists good heuristic methods, see \cite{flores2007review}. Finding a good triangulation can have huge impact on the performance of JTA which we detail in Section \ref{sec:impact_of_evidence}. A moralized and triangulated version of the graph in Figure~\ref{tikz:a} is shown Figure~\ref{tikz:b}, where no fill edges was necessary to make the graph triangulated.

A triangulated graph can always be represented as a \textit{junction tree}. A junction tree is a tree where the nodes are given by the cliques of the triangulated graph with the property that for two cliques, $C$ and $C'$, the intersection $C \cap C'$ is contained in all clique nodes on the unique path between $C$ and $C'$. The cliques of the graph in Figure~\ref{tikz:b} are given as $C_3 = \{c,d,e\}, C_2 = \{a,c,d\}$ and $C_1 = \{a,b,d\}$ where we arbitrarily designate $C_1$ as the root to obtain the rooted junction tree in Figure~\ref{tikz:c}. Now, assign each potential in \eqref{eq:pot_rep} to a clique potential for which the variables conform, e.g.
\begin{align*}
  \phi_{C_1}(x_a, x_b, x_d) &\leftarrow p(x_a \mid x_b, x_d), \\
  \phi_{C_2}(x_a, x_c, x_d) &\leftarrow p(x_c)p(x_a \mid x_c), \\
  \phi_{C_3}(x_c, x_d, x_e) &\leftarrow p(x_d \mid x_c)p(x_e \mid x_c, x_d).
\end{align*}
The clique potentials are now \textit{initialized} (we also say that the network has been initialized) and note that the clique potentials in general do not have any interpretation at this stage. We have obtained the \textit{clique potential representation}
\begin{equation}
  \label{eq:clique_pot_rep}
  p(x_a, x_b, x_c, x_d, x_e) = \phi_{C_1}(x_a, x_b, x_d)\phi_{C_2}(x_a, x_c, x_d)\phi_{C_3}(x_a, x_c, x_d).
\end{equation}
The network is said to be \textit{compiled} at this stage, i.e.\ when moralization and triangulization has been performed and the clique potential representation is obtained. In complex networks with large clique potentials it might not be feasible to even initialize the clique potentials due to lack of memory. We give an example of how to overcome this in Section \ref{sec:impact_of_evidence} by entering \textit{evidence} into the model as described in Section \ref{sec:evidence_slicing}.

Next, the message passing scheme can now be applied to the junction tree. We describe here, the Lauritzen-Spiegelhalter (LS) scheme which works as follows. Locate a \textit{leaf node}, here we choose $C_{3}$, and find the intersection, $S_{32} = C_3 \cap C_2 = \{c, d\}$, with its parent clique $C_2$. Then calculate the marginal potential $\phi_{S_{32}}(x_c, x_d) = \sum_{x_e}\phi_{C_3}(x_c, x_d, x_e)$ and perform an \textit{inward message} by setting
\[
  \phi_{C_{2}}(x_a, x_c, x_d) \leftarrow \phi_{C_{2}}(x_a, x_c, x_d)\phi_{S_{32}}(x_c, x_d)
\]
and update the leaf node as
\begin{equation}
  \label{eq:update_c2}
  \phi_{C_3}(x_c, x_d, x_e) \leftarrow \phi_{C_3}(x_c, x_d, x_e) / \phi_{S_{32}}(x_c, x_d),
\end{equation}
where $0/0:=0$. We say that $C_2$ have \textit{collected} its messages from all of its children. This procedure must be repeated until the root, $C_1$, has collected all its messages. Hence, we perform another inwards message by setting $\phi_{S_{21}}(x_c) = \sum_{x_a, x_d}\phi_{C_{2}}(x_a, x_c, x_d)$ and update:
\begin{align*}
  \phi_{C_{1}}(x_a, x_b, x_d) &\leftarrow \phi_{C_{2}}(x_a, x_c, x_d)\phi_{S_{21}}(x_c, x_d), \\
  \phi_{C_{2}}(x_a, x_c, x_d) &\leftarrow \phi_{C_{2}}(x_a, x_c, x_d) / \phi_{S_{21}}(x_c, x_d).
\end{align*}
The \textit{inward phase} terminates when the root clique potential has been normalized: 
\[
  \phi_{C_1}(x_a, x_b, x_d) \leftarrow \phi_{C_1}(x_a, x_b, x_d) / \sum_{x_a, x_b, x_d}\phi_{C_{1}}(x_a, x_b, x_d).
\]
To summarize, we have now obtained what is called the \textit{set chain representation}
\begin{align*}
  \label{eq:setchain}
  p(x_a, x_b, x_c, x_d, x_e) &= \phi_{C_1}(x_a, x_b, x_d) \phi_{C_2}(x_a, x_c, x_d) \phi_{C_3}(x_c, x_d, x_e)\\
                             &= p(x_a, x_b, x_d)p(x_a \mid x_c, x_d)p(x_c, x_d \mid x_e),
\end{align*}
where the clique potentials are now conditional probability tables. Notice especially that $\phi_{C_1}(x_a, x_b, x_d) = p(x_a, x_b, x_d)$. In the \textit{outward phase} we start by sending messages from the root by performing an \textit{outward message} by letting $\phi_{S_{12}}(x_a, x_d) = \sum_{x_b}\phi_{C_1}(x_a, x_b, x_d)$ and update:
\begin{equation}
  \phi_{C_2}(x_a, x_c, x_d) \leftarrow \phi_{C_{2}}(x_a, x_c, x_d)\phi_{S_{12}}(x_c, x_d).
\end{equation}
We say that $C_1$ has \textit{distributed} evidence to $C_2$. Notice, that $\phi_{C_2}$ is now identical to the probability distribution defined over the variables $x_a, x_c$ and $x_d$. Finally, let $\phi_{S_{23}}(x_c, x_d) = \sum_{x_a}\phi_{C_2}(x_a, x_c, x_d)$ and update $\phi_{C_3}$ as 
\[
  \phi_{C_3}(x_c, x_d, x_e) \leftarrow \phi_{C_3}(x_c, x_d, x_e) \phi_{S_{23}}(x_c, x_d).
\]
As a consequence we finally obtain the \textit{clique marginal representation}
\begin{align*}
  p(x_a, x_b, x_c, x_d, x_e) &= \frac{\phi_{C_{1}}(x_a, x_b, x_d)\phi_{C_{2}}(x_a, x_c, x_d)\phi_{C_3}(x_c, x_d, x_e)}{\phi_{S_{12}}(x_a, x_d)\phi_{S_{23}}(x_c, x_d)} \\
  &= \frac{p(x_a, x_b, x_d)p(x_a, x_c, x_d)p(x_c, x_d, x_e)}{p(x_a, x_d)p(x_c, x_d)},
\end{align*}
where all clique and separator potentials are identical to the marginal probability distribution over the variables involved. Hence we can now find $p(x_a, x_d)$ by locating a clique containing $x_a$ and $x_d$ and sum out all other variables. If we choose $C_2$ we get
\[
  p(x_a, x_d) = \sum_{x_c}\phi_{C_{2}}(x_a, x_c, x_d).
\]
Each time we multiply, divide or marginalize potentials, a number of binary operations (addition, multiplication and division) are conducted under the machinery. For a network with $41$ variables and a maximum size of the state space for each variable being $3$, \cite{lepar1998comparison} recorded a total number of $2,371,178$ binary operations. We do not intend to follow the same analysis here, but for sparse tables the number of necessary binary operations is potentially much smaller.

\subsection{Evidence and Slicing}
\label{sec:evidence_slicing}
Suppose it is known, before message passing, that $X_{E} = x^{\ast}_{E}$ for some labels $E \subset V$. We refer to $x^{\ast}_{E}$ as \textit{evidence}. Evidence can be entered into the clique potential representation \eqref{eq:clique_pot_rep} as follows. For each $v \in E$ choose an arbitrary clique, $C$, where $v\in C$ and set entries in $\phi_{C}$ that are inconsistent with $x^{\ast}_{v}$ equal to zero. The resulting clique potential is then said to be \textit{sliced}. After message passing, all queries are then conditional on $X_{E} = x^{\ast}_{E}$. Thus entering evidence leads to more zero-cells, and in a sparse setup, the resulting clique potentials will be even more sparse. After message passing, the clique potential $\phi_{C}(x_{C})$ is now equal to the conditional probability $p(x_{C\setminus E} \mid x_E^\ast)$.

It suffices to modify a single clique potential such that it is inconsistent with $v \in E$, for all $v$ as described above. However, for sparse tables it is advantageous to enter evidence in all clique potentials containing $v$ since this leads to a higher degree of sparsity. This is how evidence is handled in \jti. In fact, this is the reason why \jti is able to handle very complex networks by exploiting evidence using sparse tables from \sparta. It is also possible to enter evidence into the factorization \eqref{eq:pot_rep}. This is the key to handle complex networks that are otherwise infeasible due to lack of memory as we show in Section \ref{sec:impact_of_evidence}. This trick is related to \textit{cutset conditioning} and if one can enter evidence into variables that breaks cycles the effect can be huge \citep{pearl2013constraint}. One can exploit the law of total probability to sum out the variables that was conditioned on (since we obtain the probability of the evidence during propagation) and recover whatever probablity is of interest.

\textcolor{white}{space}
\section{An Intuitive way of Representing Sparse Tables}
\label{sec:intui}
Before describing our method for multiplication and marginalization of sparse tables, it is illuminating to describe sparse tables in a standard \proglang{R} language setup. Consider two arrays \code{f} and \code{g}:
\begin{Schunk}
\begin{Sinput}
R> dn <- function(x) setNames(lapply(x, paste0, 1:2), toupper(x))
R> d  <- c(2, 2, 2)
R> f  <- array(c(5, 4, 0, 7, 0, 9, 0, 0), d, dn(c("x", "y", "z")))
R> g  <- array(c(7, 6, 0, 6, 0, 0, 9, 0), d, dn(c("y", "z", "w")))
\end{Sinput}
\end{Schunk}
with flat layouts
\begin{multicols}{2}
\begin{Schunk}
\begin{Sinput}
R> ftable(f, row.vars = "X")
\end{Sinput}
\begin{Soutput}
   Y y1    y2   
   Z z1 z2 z1 z2
X               
x1    5  0  0  0
x2    4  9  7  0
\end{Soutput}
\begin{Sinput}
R> ftable(g, row.vars = "W")
\end{Sinput}
\begin{Soutput}
   Y y1    y2   
   Z z1 z2 z1 z2
W               
w1    7  0  6  6
w2    0  9  0  0
\end{Soutput}
\end{Schunk}
\end{multicols}
If we convert \code{f} and \code{g} to \code{data.frame} objects and exclude the cases with a value of zero we have:
\begin{Schunk}
\begin{Sinput}
R> df <- as.data.frame.table(f, stringsAsFactors=FALSE)
R> df <- df[df$Freq != 0,]
R> dg <- as.data.frame.table(g, stringsAsFactors=FALSE)
R> dg <- dg[dg$Freq != 0,]
\end{Sinput}
\end{Schunk}
\begin{multicols}{2}
\begin{Schunk}
\begin{Sinput}
R> print(df, row.names = FALSE)
\end{Sinput}
\begin{Soutput}
  X  Y  Z Freq
 x1 y1 z1    5
 x2 y1 z1    4
 x2 y2 z1    7
 x2 y1 z2    9
\end{Soutput}
\begin{Sinput}
R> print(dg, row.names = FALSE)
\end{Sinput}
\begin{Soutput}
  Y  Z  W Freq
 y1 z1 w1    7
 y2 z1 w1    6
 y2 z2 w1    6
 y1 z2 w2    9
\end{Soutput}
\end{Schunk}
\end{multicols}
which literally leaves us with two sparse tables, \code{df} and \code{dg} respectively. In order to multiply \code{df} by \code{dg} we must, by definition, determine the cases that match on the variables \code{Y,Z} that they have in common. For example, row $4$ in \code{df} must be multiplied with row $4$ in \code{dg} such that \code{(y1, z2, x2, w2)} is an element in the product with value $81$. And since the tables are sparse, no multiplication by zero will be performed. The multiplication can be performed with the following small piece of \proglang{R} code (which will be used in Section \ref{sec:benchmark} in connection with bench-marking):


\begin{Schunk}
\begin{Sinput}
R> sparse_prod <- function(df, dg){
+    S   <- setdiff(intersect(names(df), names(dg)), "Freq")
+    mrg <- merge(df, dg, by = S)
+    val <- mrg$Freq.x * mrg$Freq.y
+    mrg$val <- val
+    mrg[, setdiff(names(mrg), c("Freq.x", "Freq.y"))]
+  }
\end{Sinput}
\end{Schunk}
The \code{merge} function performs, by default, what is also called an \textit{inner join} or \textit{natural join} in SQL terminology, which is exactly how we defined table multiplication in Section \ref{sec:notation}. Multiplying \code{df} and \code{dg} yields
\begin{Schunk}
\begin{Sinput}
R> sparse_prod(df, dg)
\end{Sinput}
\begin{Soutput}
   Y  Z  X  W val
1 y1 z1 x1 w1  35
2 y1 z1 x2 w1  28
3 y1 z2 x2 w2  81
4 y2 z1 x2 w1  42
\end{Soutput}
\end{Schunk}
Marginalization is even more straightforward. Marginalizing out $X$ from \code{df} can for example be done using the built-in \proglang{R} function \texttt{aggregate} (which is also used in Section \ref{sec:benchmark} for benchmarking):
\begin{Schunk}
\begin{Sinput}
R> aggregate(Freq ~ Y + Z, data = df, FUN = sum)
\end{Sinput}
\begin{Soutput}
   Y  Z Freq
1 y1 z1    9
2 y2 z1    7
3 y1 z2    9
\end{Soutput}
\end{Schunk}
Thus, we have the necessary tools to implement JTA using sparse tables. So why should we bother redefining sparse tables and algebras on these; because of execution time and memory storage. In Section \ref{sec:benchmark} we show the effect of the effort of going beyond the \code{merge} and \code{aggregate} functions.

\section{Sparse Tables}
\label{sec:sparse_tables}
Let $T$ be a dense table with domain $I = \times_{v\in V}I_{v}$ as described in \ref{sec:notation}. Define the \textit{level set} $\mathcal{L} :=\times_{v\in V}\mathcal{L}_{v}$ where $\mathcal{L}_{v} = \{1, 2, \ldots, |I_{v}|\}$ and let $\# : I \rightarrow \mathcal{L}$ be a bijection. We define the sparse table $\tau = (\Phi, \phi)$, of $T$ as the pair where $\Phi$ is a matrix with columns given by the set of vectors in the \textit{sparse domain} $\mathcal{I} := \{\#(x) \mid T(x) \neq 0, x \in I\}$, consisting of \textit{non-zero cells} and where $\phi$ is the corresponding vector of values. Thus, a column in $\Phi$ represents a cell in $\mathcal{I}$ and is written a tuple $i = (i_{1}, i_{2}, \ldots, i_{|V|}; i_{v} \in \mathcal{L}_{v})$ which explicitly determines the ordering of the labels and hence the order of the rows in $\Phi$. The order of the columns in $\Phi$ is not important as long as it agrees with $\phi$. We denote by $\Phi[j]$ the $j'$th column of $\Phi$ and by $\phi_j$ the corresponding $j'$th value in $\phi$. The sub-matrix $\Phi_{S}$ defined over the set of labels, $S \subseteq V$, is the resulting matrix when rows corresponding to labels in $V\setminus S$ have been removed. Let $T$ be the table \code{f} from Section \ref{sec:intui}:
\begin{Schunk}
\begin{Soutput}
   Y y1    y2   
   Z z1 z2 z1 z2
X               
x1    5  0  0  0
x2    4  9  7  0
\end{Soutput}
\end{Schunk}
The domain is given by
\[
  I = \{x_1, x_2\} \times \{y_1, y_2\} \times \{z_1, z_2\}
\]
and we can choose $\#$ as the map $(x_{\ell_1}, y_{\ell_2}, z_{\ell_3}) \mapsto (\ell_1, \ell_2, \ell_3)$ for $\ell_1, \ell_2, \ell_3 \in \{1,2\}$. The non-zero cells can be identified from the table and we have $\mathcal{I} = \{(1,1,1), (2,1,1), (2,2,1), (2,1,2)\}$. Hence
\[
  \Phi = \begin{bmatrix}
    1&  2& 2& 2\\ 
    1&  1& 2& 1\\ 
    1&  1& 1& 2
  \end{bmatrix},
\]
with $\phi = (5, 4, 7, 9)$, corresponding to \code{df} in Section \ref{sec:intui}. Let $G$ be another dense table with domain $J = \times_{u\in U}J_{u}$ and sparse representation $\gamma = (\Psi, \psi)$ with sparse domain $\mathcal{J}$. We then aim at defining the sparse multiplication $\tau \otimes \gamma$ of $T \otimes G$. Let $S = V \cap U$ be the \textit{separator labels} shared between the two sparse tables $\tau$ and $\gamma$. Next, define the map $M_{S}(\Phi)$ which transform $\Phi_{S}$ into a look-up table\footnote{A lookup table is a list arranged as key-value pairs. In \proglang{R} one can think of a look-up table as a named list where the names are the keys and the values are the elements of the list.} as follows: the keys are the unique columns of $\Phi_{S}$ and the value of $M_{S}(\Phi)$ at key $k$ is the set of column indices where column $k$ can be found in $\Phi_{S}$ and hence also in $\Phi$ and is given by
\[
  M_{S}(\Phi)[k] = \{ j \in \{1, 2, \ldots, |\mathcal{I}| \} : \Phi[j] = k \}.
\]
Let $\mathcal{K}$ denote the mutual keys of $M_{S}(\Phi)$ and $M_{S}(\Psi)$. The number of columns in the matrix of the resulting product $\tau \otimes \gamma$ is then given as
\[
  N := \sum_{k\in\mathcal{K}} |M_{S}(\Phi)[k]| \cdot |M_{S}(\Psi)[k]|.
\]
This observation is crucial, since the memory storage of the sparse product can then be computed in advance. If $(\Pi, \pi)$ is the sparse product of $\tau$ and $\gamma$ we can therefore initialize $\Pi$ as a matrix with $|V| + |U\setminus V|$ rows and $N$ columns and $\pi$ as an $N-$dimensional vector. Finally, $\pi$ is given by the values $\phi_j \cdot \psi_{j'}$ for $j \in M_{S}(\Phi)[k]$ and $j' \in M_{S}(\Psi)[k]$ for all $k \in \mathcal{K}$. The procedure is formalized in Algorithm \ref{alg:mult}.

\begin{algorithm}[h!]
\caption{Multiplication of Sparse Tables}\label{alg:mult}
\begin{algorithmic}[1]
  \Procedure{}{$\tau = (\Phi, \phi)$: sparse table, $\gamma = (\Psi, \psi)$: sparse table}
  \State $S := V \cap U$
  \State $\mathcal{K}$: Mutual keys of $M_{S}(\Phi)$ and $M_{S}(\Psi)$
  \State $N := \sum_{k\in\mathcal{K}} |M_{S}(\Phi)[k]| \cdot |M_{S}(\Psi)[k]|$
  \State Initialize the matrix $\Pi$ with $|V| + |U\setminus V|$ rows and $N$ columns
  \State Initialize the vector $\pi$ of dimension $N$
  \State $l := 1$
  \For{$k \in \mathcal{K}$}
       \For{$j\in M_{S}(\Phi)[k]$ and $j' \in M_{S}(\Psi)[k]$}
       \State $\Pi[l] := (\Phi[j], \Psi_{U\setminus V}[j'])$
       \State $\pi_l  := \phi_{j} \cdot \psi_{j'}$ \label{alg:mult_pie}
       \State $l = l + 1$
      \EndFor
  \EndFor
  \State\Return $(\Pi, \pi)$    
 \EndProcedure
\end{algorithmic}
\end{algorithm}

The number of binary operations is smaller than the equivalent dense
table multiplication since every multiplication with zero is
avoided. Moreover, since we only loop over the mutual keys,
$\mathcal{K}$, the execution time will depend on the table having the
least unique keys over the separator labels. Trivially, division of
two sparse tables can be obtained by changing line \ref{alg:mult_pie}
of Algorithm \ref{alg:mult} with $\pi_l = \phi_j / \phi_j'$.

Now, let $G$ be the table \code{g} from Section \ref{sec:intui} where the domain is given by
\[
  J = \{w_1, w_2\} \times \{y_1, y_2\} \times \{z_1, z_2\}.
\]
Choose the map $(w_{\ell_1}, y_{\ell_2}, z_{\ell_3}) \mapsto (\ell_1, \ell_2, \ell_3)$ for $\ell_1, \ell_2, \ell_3 \in \{1,2\}$. Then, in summary we have the tables
\[
  \Phi = \begin{bmatrix}
    1&  2& 2& 2\\ 
    1&  1& 2& 1\\ 
    1&  1& 1& 2
  \end{bmatrix}, \quad
  \Psi = \begin{bmatrix}
    1&  2& 2 & 1\\ 
    1&  1& 2 & 2\\ 
    1&  1& 1 & 2
  \end{bmatrix},
\]
along with $\phi = (5,4,7,9)$ and $\psi = (7,6,6,9)$. The separator labels are given by $S = \{y, z\}$ and the lookup tables of $\Phi$ and $\Psi$ are then given by
\begin{align*}
  M_{S}(\Phi) &= \{ (1,1):= \{1, 2\} , (2,1): = \{3\}, (1,2):=\{4\} \} \\
  M_{S}(\Psi) &= \{ (1,1):= \{1\} , (2,1): = \{2\}, (1,2):=\{4\}, (2,2):=\{3\}\}.
\end{align*}
Above, $y$ corresponds to row two and $z$ corresponds to row three in $\Phi$. So for example, $M_{S}(\Phi)[(1,1)] = \{1,2\}$ means, that the key $(1,1)$ has value $\{1,2\}$ which means that $(1,1)$ is found in columns $1$ and $2$ in $\Phi$.

Therefore, all values $\phi_j$ for $j\in M_{S}(\Phi)[(1,1)]$ must be multiplied with all values $\psi_j$ for $j\in M_{S}(\Psi)[(1,1)]$ etc. Hence,
\[
  \Pi = \begin{bmatrix}
    1&  2& 2& 2 \\ 
    1&  1& 2& 1 \\ 
    1&  1& 1& 2 \\
    1&  1& 1& 2
  \end{bmatrix},
\]
and
\[
  \pi = (\phi_1 \cdot \psi_1, \phi_2 \cdot \psi_1, \phi_3 \cdot \psi_2, \phi_4 \cdot \psi_4) = (35,28,42,81),
\]
as expected from the result in Section \ref{sec:intui} using \code{sparse_mult}. Notice, that we save any computation with $\psi_3$ since $(2,2)$ is not a key in $M_{S}(\Phi)$.

We mention that addition and subtraction of sparse tables is more demanding since we have to reconstruct zero-cells if one of the tables
have a non-zero cell-value while the other table has a zero-cell in
the corresponding separator cell. Fortunately, these operations are not
needed in JTA.

The marginal sparse table $\tau^{\downarrow A} = (\Phi^{\downarrow A}, \phi^{\downarrow A})$ of $\tau$, corresponding to $T^{\downarrow A}$, can be calculated using the map $M_{A}(\Phi)$ and, for each key $k \in M_{A}(\Phi)$, sum the corresponding values in $\phi$. However, for massive tables the memory footprint of $M_{A}(\Phi)$ is unnecessarily large. Instead, we construct the lookup-table $H_{A}(\Phi)$ where the keys are the unique columns of $\Phi_{A}$ as was the case in $M_{A}(\Phi)$. However, the values are themselves pairs where the first element is an index to any of the column indices where the corresponding key can be found in $\Phi_{A}$. The second element is the final cell value in the marginalized table corresponding to the key. The pair corresponding to the key $k$ is therefore on the form 
\[
  H_{A}(\Phi)[k] = (j, v), \quad v = \sum_{\ell : \Phi_{A}[\ell] = k} \phi_\ell \mbox{ and } \Phi_{A}[j] = k.
\]
The value $v$ can easily be computed iteratively. The point here is, that we never have to store $\Phi_{A}$ since we can deduce all information from $\Phi$ on the fly given the row indices corresponding to $A$ in $\Phi$. The number of columns in the final matrix $\Phi^{\downarrow A}$, and hence the number of elements in $\phi^{\downarrow A}$, is given by $|H_{A}(\Phi)|$. The procedure is formalized in Algorithm \ref{alg:marg}.
\begin{algorithm}[h!]
\caption{Marginalization of Sparse Tables}\label{alg:marg}
\begin{algorithmic}[1]
  \Procedure{}{$\tau = (\Phi, \phi)$: sparse table, $A$: Set of labels}
  \State Construct $H_{A}(\Phi)$
  \State $N = |H_{A}(\Phi)|$
  \State Initialize the matrix $\Phi^{\downarrow A}$ with $|A|$ rows and $N$ columns
  \State Initialize the vector $\phi^{\downarrow A}$ of dimension $N$
  \State Let $\mathcal{K}$ be the keys of $H_{A}(\Phi)$
  \State $l:=1$
  \For{$k \in \mathcal{K}$}
  \State $(j, v) := H_{A}(\Phi)[k]$
  \State $\Phi^{\downarrow A}[l] := \Phi_{A}[j]$ \Comment deduced by picking elements from $\Phi[j]$
  \State $\phi^{\downarrow A}_l := v$
  \State $l = l+1$
  \EndFor
  \State\Return $(\Phi^{\downarrow A}, \phi^{\downarrow A})$  
 \EndProcedure
\end{algorithmic}
\end{algorithm}
Consider again the sparse table $\tau$ of $T$ and let $A = \{y, z\}$. Then, the resulting sparse marginal table have two rows corresponding to $y$ and $z$. The construction of $H_{A}(\Phi)$ is as follows. The first column in $\Phi$ is extracted and the entry corresponding to $x$ is deleted. Call the resulting vector (key) $k_1$. Now, set $H_{A}(\Phi)[k_1] = (j = 1, v = 5)$ since $\phi_1 = 5$. Extract now, the second column of $\Phi$ and let $k_2$ be the resulting key when removing the entry corresponding to $x$. Since $k_1 = k_2$ and $\phi_2 = 4$ we update $H_{A}(\Phi)[k_1] = (j = 2, v = 9)$. Proceeding this way, gives
\[
  H_{A}(\Phi) = \{(1,1) := (j = 2, v = 9), (2,1):=(j = 3, v = 7), (1,2):= (j = 4, v = 9)\}\}.
\]
Thus
\[
  \Phi^{\downarrow A} = \begin{bmatrix}
    1&  2& 1 \\ 
    1&  1& 2
  \end{bmatrix},  
\]
and $\psi^{\downarrow A} = (9, 7, 9)$. For $B \subset V$, the \textit{$i^{\ast}_{B}-$ slice} of a sparse table $\tau = (\Phi, \phi)$ is obtained by removing columns, $k$, in $\Phi$ for which $k$ does not agree with $i^{\ast}_{B}$. We leave out the formal procedure for slicing.

\subsection{Sparse tables in JTA}
Triangulation may lead to cliques for which no potential is assigned during compilation and the clique potential is then regarded as the unity table. In the dense version, this corresponds to an array filled with ones. However, the idea of sparsity will be ruined if we just adapt the dense version. For the general case we define a \textit{sparse unity table} and show how to multiply it with a non-unity table in Appendix \ref{app:sparse_unities}. In the light of message passing, these unity tables are not needed. When a message is sent to a unity, we simply replace it with the message. And as such, we avoid a lot of computational overhead.

\subsection{How to use sparta}
\label{sec:spartatables}

In order to demonstrate the use of \sparta we revisit the example from Section \ref{sec:intui} of the two (dense) tables \code{f} and \code{g} with mutual variables, \code{Y} and \code{Z}
\begin{multicols}{2}
\begin{Schunk}
\begin{Sinput}
R> ftable(f, row.vars = "X")
\end{Sinput}
\begin{Soutput}
   Y y1    y2   
   Z z1 z2 z1 z2
X               
x1    5  0  0  0
x2    4  9  7  0
\end{Soutput}
\begin{Sinput}
R> ftable(g, row.vars = "W")
\end{Sinput}
\begin{Soutput}
   Y y1    y2   
   Z z1 z2 z1 z2
W               
w1    7  0  6  6
w2    0  9  0  0
\end{Soutput}
\end{Schunk}
\end{multicols}
We can convert these to their equivalent \code{sparta} versions as
\begin{Schunk}
\begin{Sinput}
R> sf <- as_sparta(f); sg <- as_sparta(g)
\end{Sinput}
\end{Schunk}
Printing \code{sf}
\begin{Schunk}
\begin{Sinput}
R> print.default(sf) 
\end{Sinput}
\begin{Soutput}
  [,1] [,2] [,3] [,4]
X    1    2    2    2
Y    1    1    2    1
Z    1    1    1    2
attr(,"vals")
[1] 5 4 7 9
attr(,"dim_names")
attr(,"dim_names")$X
[1] "x1" "x2"

attr(,"dim_names")$Y
[1] "y1" "y2"

attr(,"dim_names")$Z
[1] "z1" "z2"

attr(,"class")
[1] "sparta" "matrix"
\end{Soutput}
\end{Schunk}
where we use the default print method to look under the hood. The columns are the cells in the sparse matrix and the \code{vals} attribute are the corresponding values which can be extracted with the \code{vals} function. Furthermore, domain resides in the \code{dim_names} attribute which can also be extracted using the \code{dim_names} function. From the output, we see that (\code{x2}, \code{y2}, \code{z1}) has a value of $2$. Using the regular print method prettifies things:
\begin{Schunk}
\begin{Sinput}
R> print(sf)
\end{Sinput}
\begin{Soutput}
  X Y Z val
1 1 1 1   5
2 2 1 1   4
3 2 2 1   7
4 2 1 2   9
\end{Soutput}
\end{Schunk}
where row $i$ corresponds to column $i$ in the sparse matrix. We settled for this print method because printing column wise leads to unwanted formatting when the values are decimal numbers.
Consider the cell (2,1,1). The corresponding named cell is then
\begin{Schunk}
\begin{Sinput}
R> get_cell_name(sf, sf[, 2L])
\end{Sinput}
\begin{Soutput}
   X    Y    Z 
"x2" "y1" "z1" 
\end{Soutput}
\end{Schunk}
where \code{sf[, 2L]} is the second column (row in the output) of \code{sf} which is $(2,1,1)$. The product of \code{sf} and \code{sg} is
\begin{Schunk}
\begin{Sinput}
R> mfg <- mult(sf, sg); mfg
\end{Sinput}
\begin{Soutput}
  X Y Z W val
1 2 1 2 2  81
2 2 2 1 1  42
3 1 1 1 1  35
4 2 1 1 1  28
\end{Soutput}
\end{Schunk}
The equivalent dense table has $2^4=16$ entries. However, \code{mfg} stores $20$ values after all, $16$ of which are information about the cells. That is, there is some overhead storing the information about the cells, see Section \ref{sec:when}. Converting \code{sf} into a conditional probability table (CPT) with conditioning variable \code{Z}:
\begin{Schunk}
\begin{Sinput}
R> sf_cpt <- as_cpt(sf, y = "Z"); sf_cpt
\end{Sinput}
\begin{Soutput}
  X Y Z   val
1 1 1 1 0.312
2 2 1 1 0.250
3 2 2 1 0.438
4 2 1 2 1.000
\end{Soutput}
\end{Schunk}
We can further slice \code{sf_cpt} to obtain a probability distribution conditioned on \code{Z = z1}
\begin{Schunk}
\begin{Sinput}
R> slice(sf_cpt, s = c(Z = "z1"))
\end{Sinput}
\begin{Soutput}
  X Y Z   val
1 1 1 1 0.312
2 2 1 1 0.250
3 2 2 1 0.438
\end{Soutput}
\end{Schunk}
Marginalizing out \code{Y} in \code{sg} yields
\begin{Schunk}
\begin{Sinput}
R> marg(sg, y = c("Y"))
\end{Sinput}
\begin{Soutput}
  Z W val
1 2 2   9
2 2 1   6
3 1 1  13
\end{Soutput}
\end{Schunk}
which is in correspondence with the example in Section \ref{sec:sparse_tables}. Finally, we mention that a sparse table can be created using the constructor \code{sparta_struct} which can be necessary to use if the corresponding dense table is too large to have in memory. A table with relevant methods in \sparta is given in Table \ref{tab:sparta_funcs}.


\begin{table}[h!]
\centering
\begin{tabular}{@{}ll@{}}
\midrule
Function name              & Description                                          \\ \midrule
\code{as_<sparta>}         & Convert \code{array}-like object to a \code{sparta}  \\
\code{as_<array/df/cpt>}   & Convert \code{sparta} object to an \code{array/data.frame/CPT} \\
\code{sparta_struct}       & Constructor for \code{sparta} objects                \\
\code{mult, div, marg, slice} & Multiply/divide/marginalize/slice                 \\
\code{normalize}           & Normalize (the values of the result sum to one)      \\
\code{get_val}             & Extract the value for a specific named cell           \\
\code{get_cell_name}       & Extract the named cell                                \\
\code{get_values}          & Extract the values                                   \\
\code{dim_names}           & Extract the domain                                   \\
\code{names}               & Extract the variable names                           \\    
\code{max/min}             & The maximum/minimum value                            \\
\code{which_<max/min>_cell}& The column index referring to the max/min value      \\
\code{which_<max/min>_idx} & The configuration corresponding to the max/min value \\    
\code{sum}                 & Sum the values                                       \\
\code{equiv}               & Test if two tables are identical up to permutations of the columns \\ \midrule
\end{tabular}
\caption{Description of essential functions from the \sparta package.}
\label{tab:sparta_funcs}
\end{table}

\newpage

\subsection{When to Use Sparta}
\label{sec:when}
\begin{figure}[t] \centering
  \includegraphics[width=1\textwidth]{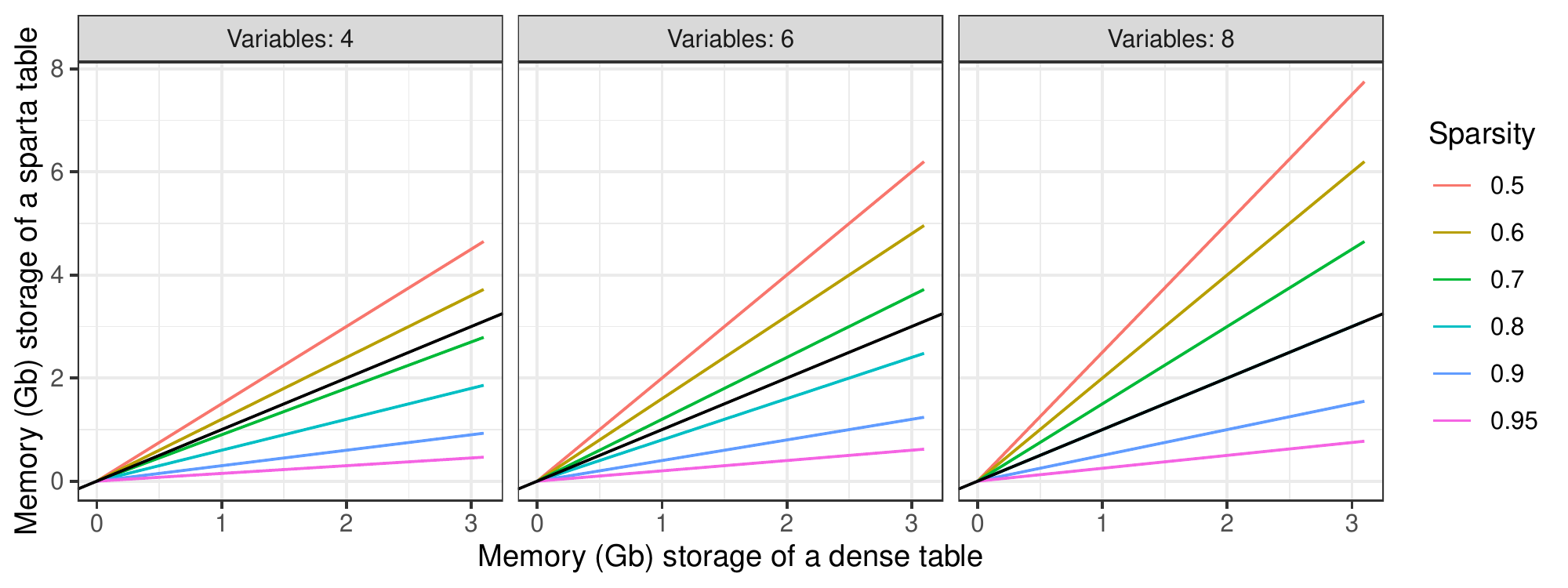}
  \caption{The black identity line indicates the number of gigabytes
    needed to store the dense table with $x$ elements. The colored
    lines indicates the number of gigabytes required to store the
    equivalent \sparta object with the respective number of variables
    and sparsity.}
  \label{fig:when}
\end{figure}

As shown in Section~\ref{sec:spartatables}, there is an overhead of
storing the information in a \sparta object. A dense array with $x$
elements takes up $8x$ bytes plus some negligible memory of storing
the variable names and etc. On the contrary, a \sparta object with
$y < x$ elements takes up $y(4k + 8)$ bytes, where $k$ is the number
of variables (these can be stored as integers and hence only requires
$4$ bytes each). In Figure~\ref{fig:when}, we have plotted this
relation for $k = 4, 6$ and $8$, and different levels of
sparsity. That is, a sparsity of $1/2$ implies that $y = x/2$. The
black identity line indicates the number of gigabytes needed to store
the dense table with $x$ elements. The size of the state spaces of the
variables are implicitly reflected by the memory needed to store the
dense table. The more memory needed, the bigger state space of the
variables. However, the more variables and the bigger state space of
these variables will intuitively result in a more sparse table, making
\sparta efficient even for several variables. 

The take away message from Figure \ref{fig:when} is that when the
state space of the variables and the sparsity increases the benefit of
storing the tables using \sparta will outweigh the overhead of storing
the additional information.

In connection to JTA, \sparta is favorable when cliques with many variables
imply a high degree of sparsity. In particular, this is often the case for tables in a Bayesian network representing a pedigree. In this case, cliques tend to be small but the state space of the variables can be arbitrarily large due to the large amount of DNA information for each member of the pedigree. In Section \ref{sec:mrf} we fit a decomposable MRF to real data and show that the sparsity of the clique potentials is much favorable towards \sparta.

\section{Usecases of jti and sparta}
\label{sec:usecases}

\begin{figure}[!h] \centering
  \includegraphics[width=0.3\textwidth]{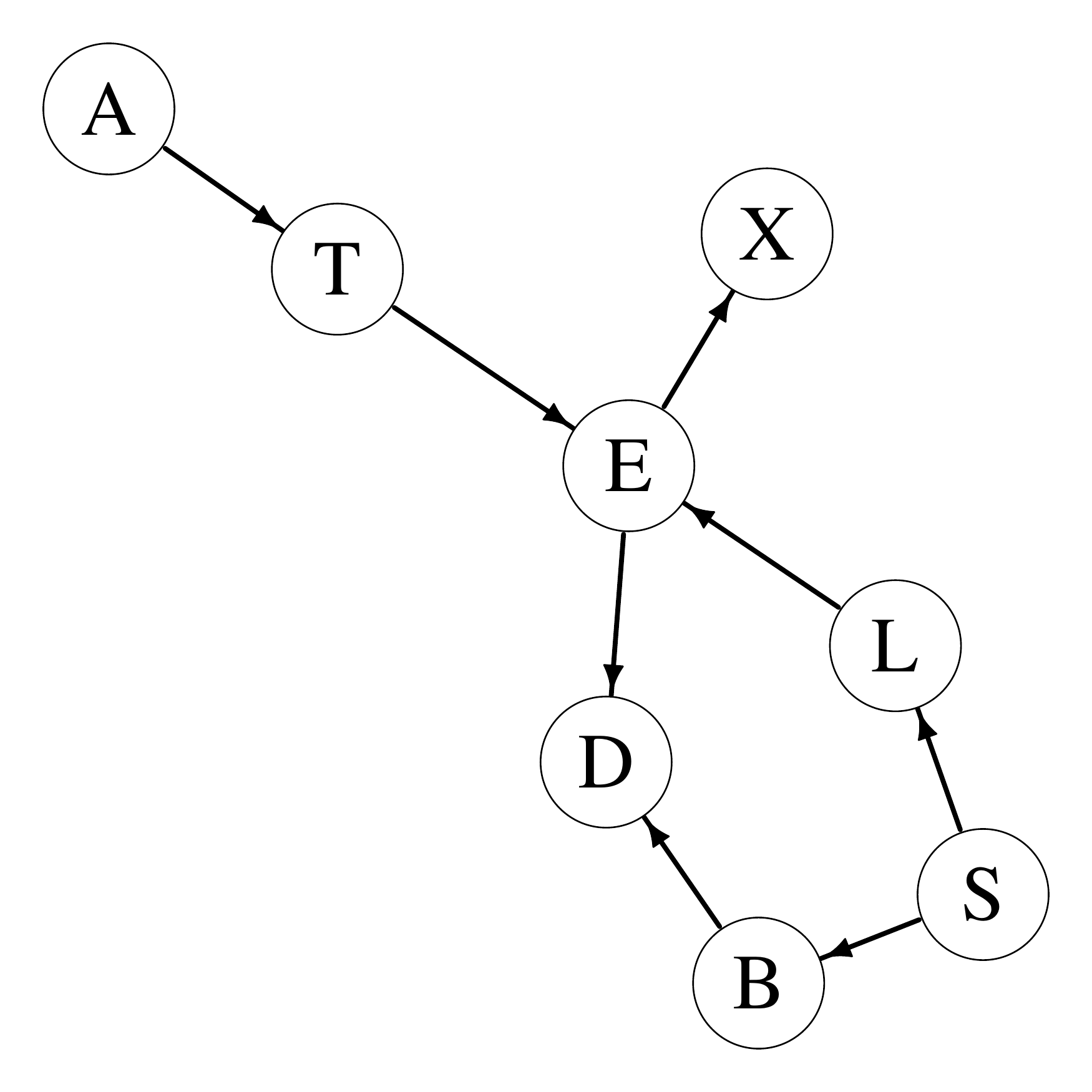}
  \caption{The DAG for the \texttt{asia} network.}
  \label{fig:asia}
\end{figure}

In \jti there are two ways of specifying a Bayesian network. Either by a list of CPTs or a dataset together with a DAG. In the latter case, the CPTs are found using maximum likelihood estimates. Here, we describe how to use \jti using the classic Bayesian network \code{asia} \citep{lauritzen1988local} where the corresponding CPTs is part of \jti. The network represents a simplified model to help diagnose the patients arriving at a respiratory clinic. A history of smoking has a direct influence on both whether or not a patient has bronchitis and whether or not a patient has lung cancer. Both lung cancer and bronchitis can result in dyspnoea. An x-ray result depends on the presence of either tuberculosis or lung cancer. Finally, a visit to Asia influence the probability of having tuberculosis. The DAG is depicted in Figure \ref{fig:asia}.

We use the version of \texttt{asia} called \texttt{asia2} which is a list of CPTs and which is shipped with \jti. The first step is to call \texttt{cpt\_list} for some initial checks and conversion to \sparta tables:
\begin{Schunk}
\begin{Sinput}
R> cl <- cpt_list(asia2)
R> cl
\end{Sinput}
\begin{Soutput}
 List of CPTs 
 -------------------------
  P( asia )
  P( tub | asia )
  P( smoke )
  P( lung | smoke )
  P( bronc | smoke )
  P( either | lung, tub )
  P( xray | either )
  P( dysp | bronc, either )

  <cpt_list, list> 
 -------------------------
\end{Soutput}
\end{Schunk}
From the output we see the inferred CPTs corresponds to Figure \ref{fig:asia} giving rise to a factorization in the same way as in (\ref{eq:pot_rep}). The network is now ready for compilation which involves moralization and triangulation. In \jti there are four different choices for triangulation which are all based on the \textit{elimination game} algorithm, see \citep{flores2007review}. One of the most well-known heuristics is \code{min_fill} which tries to minimize the number of fill edges. Evidence can be entered either at compile stage or just before message passing begins. It is always advisable to enter evidence at compile stage since we know from Section \ref{sec:evidence_slicing} that this reduces the number of non-zero elements in the CPTs and hence the memory footprint and execution time. A good strategy might be to locate one or more of the largest cliques and enter evidence on the nodes contained in these. We can investigate the cliques and their state spaces prior to compilation by triangulating the graph as follows:
\begin{Schunk}
\begin{Sinput}
R> tri <- triangulate(cl, tri = "min_fill")
\end{Sinput}
\end{Schunk}
The \code{tri} object is a list containing the triangulated graph, \code{new_graph} as a matrix, a list of \code{fill_edges}, the \code{cliques} and the size of the dense \code{statespace} of each clique. In Section \ref{sec:impact_of_evidence} we show how to exploit information about the cliques with largest state space in order to compile and propagate in a network that is otherwise infeasible on a 'standard' laptop. Now, let for example \code{tub = yes} be the evidence indicating that a given person has tuberculosis. The compiled network is then constructed as

\begin{figure}[h!] \centering
  \includegraphics[width=0.3\textwidth]{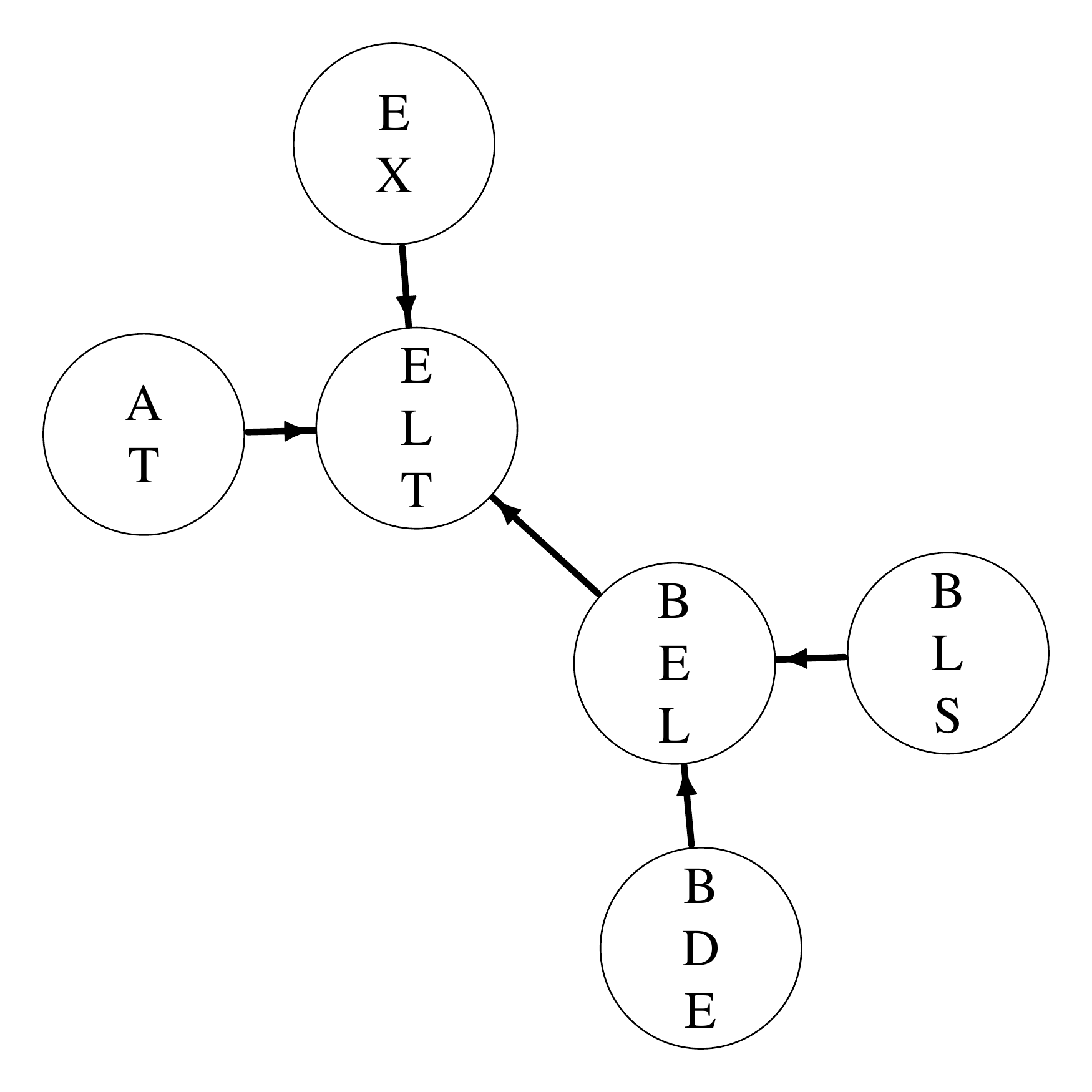}
  \caption{A junction tree for the asia network.}
  \label{fig:asia_junc}
\end{figure}

\begin{Schunk}
\begin{Sinput}
R> cp <- compile(cl, evidence = c(tub = "yes"), tri = "min_fill")
R> cp
\end{Sinput}
\begin{Soutput}
 Compiled network 
 ------------------------- 
  Nodes: 8 
  Cliques: 6 
   - max: 3 
   - min: 2 
   - avg: 2.67
  Evidence:
   - tub: yes
  <charge, list> 
 -------------------------
\end{Soutput}
\end{Schunk}
The cliques can be extracted from the compiled object with \code{get_cliques(cp)}. The compiled object can now be entered into the message passing procedure as:
\begin{Schunk}
\begin{Sinput}
R> j <- jt(cp)
\end{Sinput}
\end{Schunk}
The junction tree can be visualized by plotting the object as \code{plot(j)}, see Figure \ref{fig:asia_junc}. Finally, we can calculate the probability of a given person having a positive x-ray result, \code{xray = yes}, given that the person has tuberculosis as
\begin{Schunk}
\begin{Sinput}
R> query_belief(j, nodes = "xray")
\end{Sinput}
\begin{Soutput}
$xray
xray
 yes   no 
0.98 0.02 
\end{Soutput}
\end{Schunk}

Thus, given that a person has tuberculosis, the probability of observing a positive x-ray result is 0.98. The probability of observing positive x-ray result given that \code{tub = no} can be calculated accordingly and equals 0.1012. Joint queries can be calculated by specifying \code{type = "joint"} in \code{query_belief}.

\newpage

\subsection{Inference in Decomposable Markov Random Fields}
\label{sec:mrf}
\begin{figure}[!h] \centering
  \includegraphics[width=0.5\textwidth]{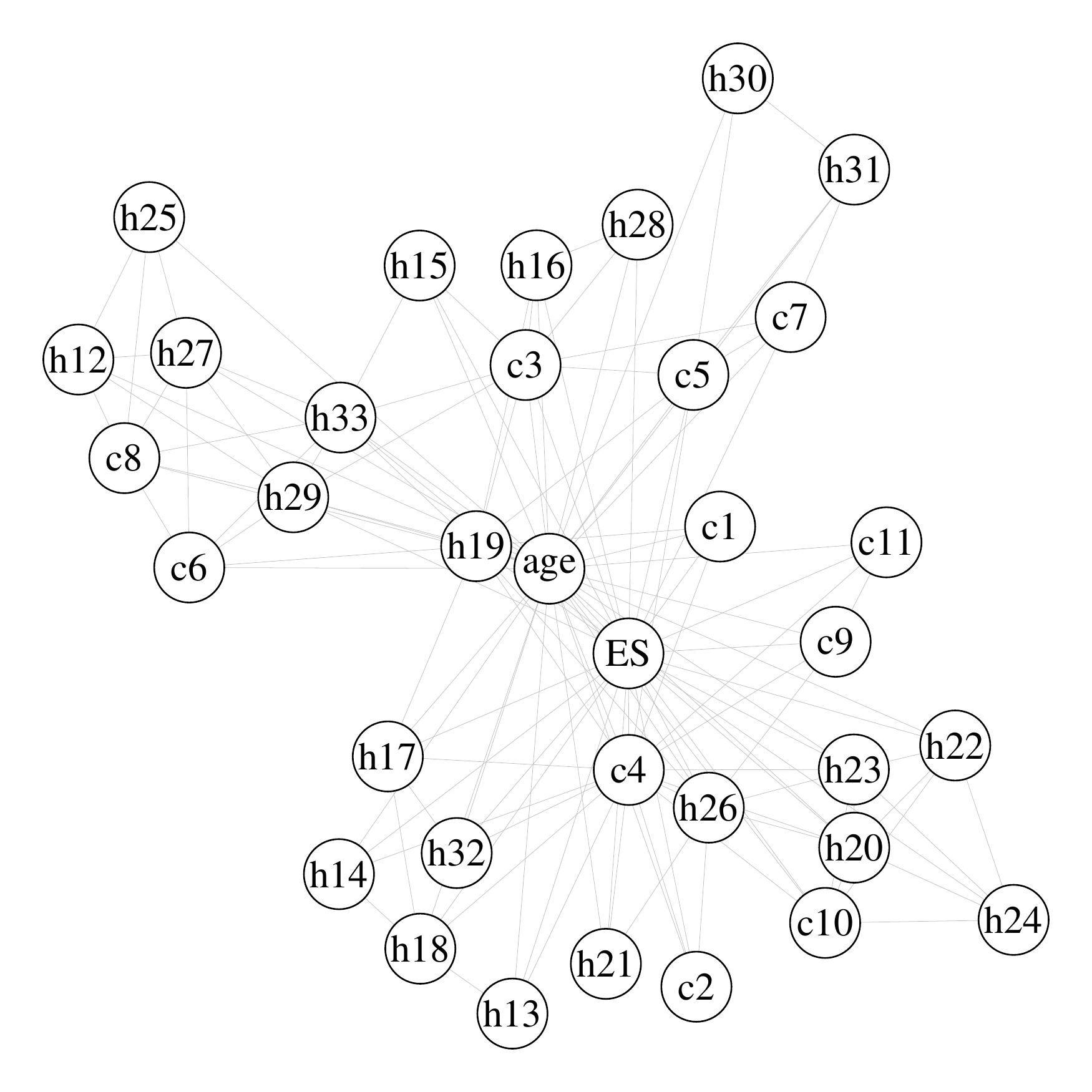}
  \caption{The fitted decomposable MRF for the derma data set.}
  \label{fig:derma}
\end{figure}

In this section we illustrate the gain of using \sparta in connection with the public-domain \texttt{derma} data set (part of the \ess package) originally obtained from the UCI Machine Learning Repository \citep{Dua:2019}. The data set consists of $358$ observations, $35$ variables of which there is one class variable called \texttt{ES}. The class variable has six states; each representing a skin disease. The remaining $34$ clinical variables (all with $4$ states except \texttt{age} which has been discretized into six bins) are used to predict the skin disease. We first fit a decomposable MRF
\begin{Schunk}
\begin{Sinput}
R> g <- ess::fit_graph(derma, q = 1.5, sparse_qic = TRUE)
\end{Sinput}
\end{Schunk}
The fitted graph is plotted in Figure \ref{fig:derma}, where we just notice, that it is a rather complex graph with large cliques. Hence, for such a small data set and a graph with big cliques, the chance of sparse clique potentials is significant.
We can now convert \code{g} into a Bayesian network simply by entering into the \code{cpt_list} function and from there compile and propagate the network:
\begin{Schunk}
\begin{Sinput}
R> cl <- jti::cpt_list(derma,  ess::as_igraph(g))
R> j  <- jti::jt(jti::compile(cl, root_node = "ES"), propagate = "no")
\end{Sinput}
\end{Schunk}
The argument \code{root_node = "ES"} forces \code{ES} into the root clique. Hence, if we are interested in queries about \code{ES} we only need to conduct the inwards message passing, collect, since the root clique potential is identical to the probability distribution over the variables in the root clique. The argument \code{propagate = "no"} means that messages passing should be postponed. By doing so, initialization is only performed once and the junction tree can be exploited as many times as needed with different evidence.

The largest clique of \code{j} has 6 variables and the mean sparsity of the clique potential is given by
\begin{Schunk}
\begin{Sinput}
R> mean(sapply(j$charge$C, sparta::sparsity))
\end{Sinput}
\begin{Soutput}
[1] 0.9233838
\end{Soutput}
\end{Schunk}
showing that the tables are extremely sparse. The junction tree can now be used for e.g. classification. Consider the first observation in \code{derma}
\begin{Schunk}
\begin{Sinput}
R> z <- unlist(derma[1, -ncol(derma)])
\end{Sinput}
\end{Schunk}
and update the junction tree with evidence corresponding to \code{z} and perform inwards message passing
\begin{Schunk}
\begin{Sinput}
R> jz <- jti::propagate(jti::set_evidence(j, z), prop = "collect")
\end{Sinput}
\end{Schunk}
The probability distribution of the class variable given the evidence \code{z} is then
\begin{Schunk}
\begin{Sinput}
R> q <- jti::query_belief(jz, nodes = "ES", type = "marginal")[[1]]
R> q[which.max(q)]
\end{Sinput}
\begin{Soutput}
seboreic dermatitis 
                  1 
\end{Soutput}
\end{Schunk}
Hence, we classify \code{z} as an observation from class seboreic dermatitis.

\subsection{The Impact of Evidence}
\label{sec:impact_of_evidence}
The Bayesian Network \code{Link} \citep{jensen1999885} is a large Bayesian network with $724$ nodes and $1,125$ arcs. The network has been used for linkage analysis where, previously, only approximate methods has been applied to make inference in the network.

We have verified that \sparta generates a large amount of overhead in the CPTs for the \code{Link} network. In fact, \jti uses approximately $16$ times more memory to store the clique potentials than the dense version. However, in the following we show how \sparta leverages from entering evidence and hence reducing the CPTs.

We picked a handful of the most comprehensive software packages, across different programming languages, for belief propagation and test if they were able to handle \code{Link} on a ``standard'' laptop machine. We used the \proglang{R} package \gRain, the \proglang{R} package \pkg{BayesNetBP}, the \proglang{Python} package \pkg{pyArum} \citep{gonzales2017agrum}, the \proglang{Julia} package \pkg{BayesNet} \citep{bayesnets}. Interestingly, all of these failed due to lack of computer memory on the first authors machine\footnote{See Section \ref{sec:comp_details} for details about this machine.} while \jti succeeded. The \pkg{pyAgrum} package is a high-level interface to the extremely efficient \proglang{C++} library \pkg{aGrum} \citep{gonzales2017agrum}. In \cite{gonzales2017agrum} the authors of \pkg{pyAgrum} themselves failed to propagate in \code{Link} on a computer with $32$Gb memory. An investigation revealed that exploiting the triangulation found by \jti, \pkg{gRain} is now also able to make inference in \code{Link}. We have not tried this with the other packages mentioned but conjecture, that the other packages would also be able to handle the \code{Link} network. Interestingly, this triangulation was the result of one of the most well-known heuristics, namely \code{min_fill}. The reason that triangulation have such a big impact is, that it determines the resulting cliques and hence the size of the clique potentials. If the cliques are too big, and hence the number of elements in the state space is large, it may not be possible to initialize the clique potentials due to lack of memory.

We now compare the cliques with larges state spaces resulting from the two heuristics \code{min_fill} and \code{min_nei} (a variant of \code{min_fill} which also tries to minimize the number of fill-in edges). We have extracted the \code{Link} network from \url{https://www.bnlearn.com/bnrepository/} as a list of CPTs with information about child and parent-node relations. In \jti we just need a list of CPTs which we extract as
\begin{Schunk}
\begin{Sinput}
R> cpts <- bnfit_to_cpts(Link)
\end{Sinput}
\end{Schunk}

Next, we convert these CPTs to their \sparta equivalent using \code{cpt_list} while deducing the underlying network, conducting some sanity checks and triangulate with the two heuristics
\begin{Schunk}
\begin{Sinput}
R> cl <- cpt_list(cpts)
R> tri_min_fill <- triangulate(cl, tri = "min_fill")
R> tri_min_nei  <- triangulate(cl, tri = "min_nei")
\end{Sinput}
\end{Schunk}
The size of the five largest clique state spaces can then be extracted as
\begin{Schunk}
\begin{Sinput}
R> sp_min_fill <- sort(tri_min_fill$statespace, decreasing = TRUE)[1:5]
R> sp_min_nei  <- sort(tri_min_nei$statespace, decreasing = TRUE)[1:5]
\end{Sinput}
\end{Schunk}
and the memory usage, for dense tables, in gigabytes of allocating memory for each of these are
\begin{Schunk}
\begin{Sinput}
R> round(sum(sp_min_fill) / 1e9*8, 2)
\end{Sinput}
\begin{Soutput}
[1] 0.2
\end{Soutput}
\begin{Sinput}
R> round(sum(sp_min_nei) / 1e9*8, 2)
\end{Sinput}
\begin{Soutput}
[1] 26.98
\end{Soutput}
\end{Schunk}

\newpage

This indicates, that any software for message passing should be able to handle the link network on a standard laptop using the \code{min_fill} heuristic from \jti, but also that \code{min_nei} will fail. 

We imagine that there is no triangulation of \code{Link}, such that any software packages, including \jti, are able to compile and propagate in \code{Link} on a standard laptop. In what follows we show, that by exploiting evidence and sparsity, we can overcome the initialization step by using \jti. We use the \code{min_nei} heuristic since this was the worst, locate one of the cliques with largest state space and insert evidence on $10$ nodes.

\begin{Schunk}
\begin{Sinput}
R> idx_max  <- which.max(tri_min_nei$statespace)
R> max_vars <- tri_min_nei$cliques[[idx_max]]
R> max_dim  <- jti::dim_names(cl)[max_vars[1:10]]
R> e        <- sapply(max_dim, `[[`, 1L) # get first state of all variables
R> cp       <- jti::compile(cl, e, tri = "min_nei")
R> j        <- jti::jt(cp)
\end{Sinput}
\end{Schunk}
The memory size (Gb), in the dense sense, in the resulting five largest clique potentials is
\begin{Schunk}
\begin{Sinput}
R> dense_sp <- function(x) prod(sapply(sparta::dim_names(x),length))
R> round(sort(sum(sapply(j$charge$C, dense_sp) / 1e9*8), TRUE), 2)
\end{Sinput}
\begin{Soutput}
[1] 0.06
\end{Soutput}
\end{Schunk}
which is negligible compared to 26.98 Gbs. The memory size (Gb) of the sparse tables is
\begin{Schunk}
\begin{Sinput}
R> round(sum(sort(sapply(j$charge$C, ncol), TRUE)[1:5]) / 1e9*8, 2)
\end{Sinput}
\begin{Soutput}
[1] 0.02
\end{Soutput}
\end{Schunk}

\section{Time and Memory Tradeoff in Sparta}
\label{sec:benchmark}
We investigate the tradeoff between memory allocation and execution
time for multiplication and marginalization on \sparta objects. If
\sparta objects do not perform reasonably well for small and medium
sized tables their usage is limited in real usecases.

Thus, we compare three functions for multiplication and three
functions for marginalization: 1) The \code{mult} and \code{marg}
functions from \sparta, 2) the \code{tabMult} and \code{tabMarg}
functions from \gRbase and 3) the \code{sparse_prod} and \code{aggregate} functions from Section \ref{sec:intui}. In the latter case we just refer to \pkg{R} as the package.

\begin{figure}[!ht] \centering
  \includegraphics[width=\textwidth]{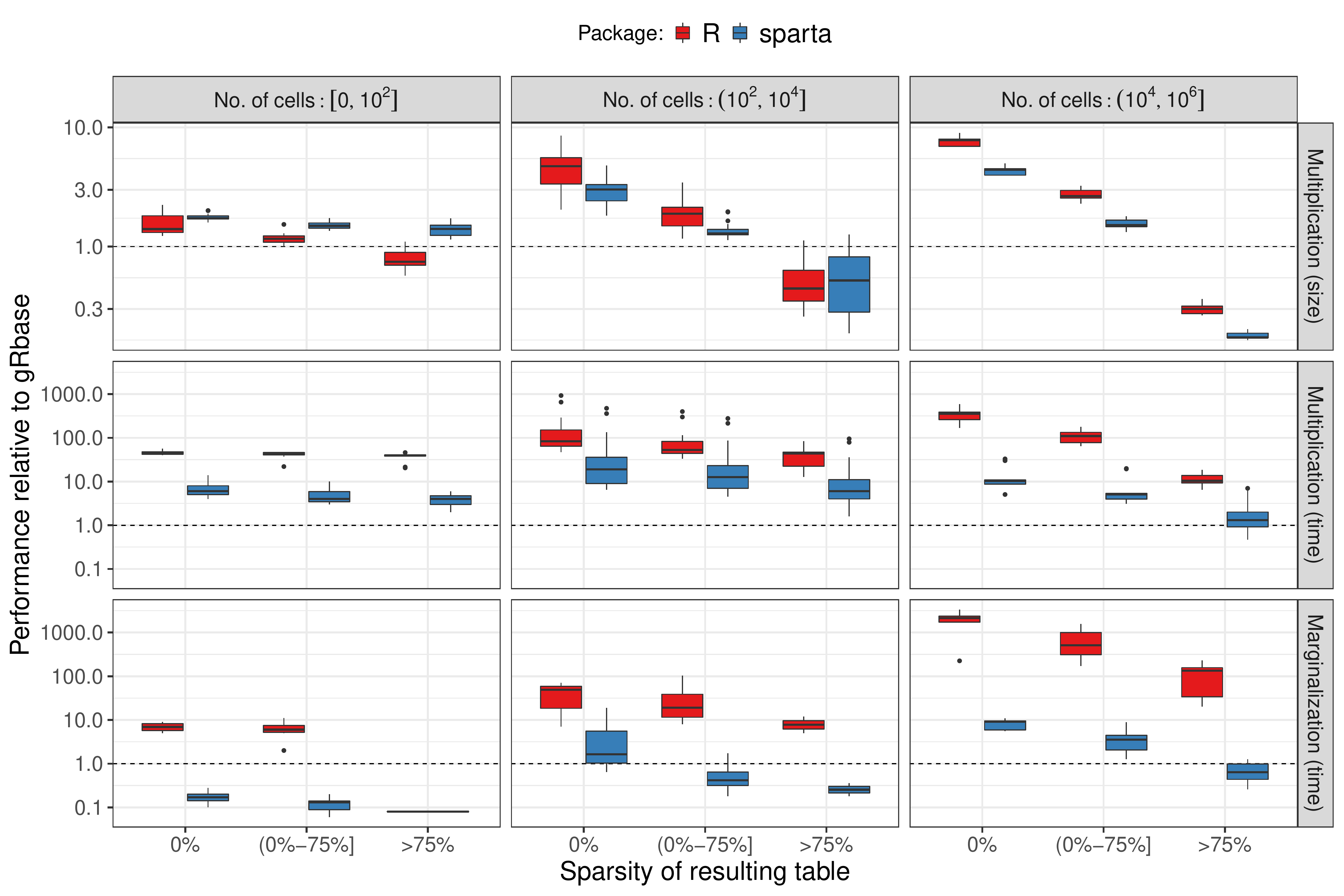}
  \caption{\label{fig:relative}Relative comparison of \pkg{R}
    (\texttt{sparse\_prod} and \texttt{aggregate}) and \sparta
    (\texttt{mult} and \texttt{marg}) to \gRbase (\texttt{tabMult} and
    \texttt{tabMarg}) in terms of memory usage and timing. The top row
    shows the relative memory usage for multiplication, whereas the
    two lower rows shows the relative timing for multiplication and
    marginalizaion, respectively.}
\end{figure}

We randomly generate pairs of tables such that the number of cells in
the product of the two tables does not exceed $10^6$. We varied the
sparsity of the product of the tables; $0\%$ sparsity (dense tables),
 $(1-75]\%$ sparsity, and  $(75, 99]\%$ sparsity. For each pair of tables, we multiply them together and record the memory usage (in megabytes) of the product and the execution time (in seconds). As \gRbase is the standard and most
mature package for graphical models in \proglang{R}, the performance
comparisons are relative to that of \gRbase.  Hence, in
Figure~\ref{fig:relative}, \gRbase performs better for tables with the
relative scores above one (indicated by horizontal dashed lines),
whereas values below one show cases where the alternative approaches
are better. The comparisons are plotted for different ranges of table
sizes (panels) and sparsity of the resulting table (first axis).

\paragraph{Multiplication (size):}

The first row of Figure~\ref{fig:relative} describes the size of the
table resulting from multiplying two tables. It can be seen that
\pkg{R} consistently produces tables of smaller sizes than \sparta for
very small tables with $100$ ($10^2$) cells. For tables with more than
$100$ cells, \sparta consistently produces smaller tables except for a
single case. Increasing the degree of sparsity leads to reduced object
sizes for both \pkg{R} and \sparta, and for tables with more than
$75\%$ sparsity, \sparta outperforms both \pkg{R} and \grbase expect
in the first two panels with small tables.

\paragraph{Multiplication (time):}

The second row of Figure~\ref{fig:relative} describes the computing
time for multiplying two tables. Clearly \sparta outperforms \pkg{R}
by orders of magnitude. For larger tables, (the two rightmost
columns), there is also a clear effect of the degree of sparsity on
the computing time.

\paragraph{Marginalisation (time):}

The third row of Figure~\ref{fig:relative} describes the computing
time for marginalising a table. When the degree of sparsity increases,
the computing time decreases. In the comparison between \gRbase and
\sparta, we see that the marginalization implementation of \sparta is
competitive to that of \gRbase. For tables with $10,000$ ($10^4$) or
fewer cells it is faster irrelevant of the sparsity except in a single
case. For 75\% sparsity \sparta's marginalization is consistently
faster.

This small, and non-exhaustive, benchmark study indicates, that our
proposed method for table multiplication and marginalization performs
well for small, medium and large tables. However, our real interest is
in the performance on massive tables which is impossible to benchmark
in this paper due to the increased running time and memory usage of
\gRbase and \pkg{R}.

\newpage

\section{Summary}
We have presented a novel method for multiplication and
marginalization of sparse tables. The method is implemented in the
\proglang{R} package \sparta. However the method is generic and we
have provided detailed pseudo algorithms facilitating the extension to
other languages. In addition we presented the companion package \jti
to illustrate some of the advantages of \sparta in connection to the
Junction Tree Algorithm. We hope to explore the benefit from the
\proglang{C} API for working with external pointers in order to reduce
the memory usage for \sparta objects in the future. 

The memory footprint of the clique potentials can become prohibitively large when the size of the cliques are large This may not be true in general for sparse tables. As a matter of fact, it may be optimal to have large cliques if they are very sparse and/or if it is common to observe variables in such a clique. Finally, we hope to explore pedigree networks that can potentially benefit immensely from \sparta in the future. 

\section*{Computational details}
\label{sec:comp_details}
In addition to the packages already mentioned, the following \proglang{R} packages was used to make the benchmark results

\begin{itemize}
\item \code{dplyr} version $1.0.6$
\item \code{glue} version $1.4.2$
\item \code{tictoc} version $1.0$
\item \code{ggplot2} version $3.3.3$
\end{itemize}

We used \proglang{R} version $4.1.0$. All computations were carried out on a $64$-bit Linux computer with Ubuntu $20.04.2$ and Intel(R) Core(TM) i7-6600U CPU 2.60GHz LTS. The machine has approximately $6$Gb of free memory for use in calculations.

\appendix
\section{Sparse unity tables}
\label{app:sparse_unities}
We define a \textit{sparse unity table}, $\mu$, over the variables $U$ to equal the set of level sets $\{\mathcal{L}_{u}\}_{u\in U}$ induced by the domain $J$. Multiplying a sparse table, $\tau = (\Phi, \phi)$, defined over the set of labels $V$ with sparse domain $\mathcal{I}$ and the sparse unity table, $\mu$, can be performed as follows. Define $R = U\setminus V$ and  $\mathcal{L}_{R} = \times_{r\in R}\mathcal{L}_{r}$ and initialize the product matrix $\Pi$ with $|V| + |R|$ rows and $|\mathcal{I}| \cdot |\mathcal{L}_{R}|$ columns. In other words, $\Pi$ has $|\mathcal{L}_{R}|$ times more columns than $\Phi$. For $j = 1, 2, \ldots, |\mathcal{I}|$ and $i_{R} \in \mathcal{L}_{R}$ append $(\Phi[j], i_{R})$ to $\Pi$ and set the corresponding value in $\pi$ to $\phi_j$. That is, we multiply a sparse table with a sparse unity table on the fly. When $|I| \gg |\mathcal{I}|$ this approach greatly reduces the number of binary operations compared to the dense case; in fact, there are no binary operations. The procedure is summarized in Algorithm \ref{alg:mult_unity}. However, dividing $\mu$ with $\tau$, one must change line \ref{alg:mult_unity_pie} in Algorithm \ref{alg:mult_unity} with $\pi_l := 1 / \phi_j$.

\begin{algorithm}[h!]
\caption{Multiplication of a Sparse Table with a Sparse Unity Table}\label{alg:mult_unity}
\begin{algorithmic}[1]
  \Procedure{}{$\tau = (\Phi, \phi)$: sparse table, $\mu$: sparse unity table}
  \State $R := U \setminus V$
  \State Set $N = |\mathcal{I}| \cdot |\mathcal{L}_{R}|$
  \State Initialize the matrix $\Pi$ with $|V| + |R|$ rows and $N$ columns
  \State Initialize the vector $\pi$ of dimension $N$
  \State $l := 1$
  \For{$j = 1, 2, \ldots, |\mathcal{I}|$}
       \For{$i_{R} \in \mathcal{L}_{R}$}
          \State $\pi[l] := (\Phi[j], i_{R})$
          \State $\pi_l := \phi_j$ \label{alg:mult_unity_pie}
          \State $l = l + 1$
      \EndFor
  \EndFor
  \State\Return $(\Pi, \pi)$    
 \EndProcedure
\end{algorithmic}
\end{algorithm}

\end{document}